\documentclass[aps,prc,preprint,showpacs,floatfix]{revtex4-1}
\usepackage[dvips]{color}
\usepackage{graphicx}
\usepackage{bm}

\begin{document}

\title{Spontaneous fission of the superheavy nucleus $^{286}$Fl}

\author{D. N. Poenaru$^*$ and R. A. Gherghescu}
\email[]{poenaru@fias.uni-frankfurt.de}
\affiliation{
Horia Hulubei National Institute of Physics and Nuclear
Engineering (IFIN-HH), \\P.O. Box MG-6, RO-077125 Bucharest-Magurele,
Romania and\\ Frankfurt Institute for Advanced Studies, Johann Wolfgang
Goethe University, Ruth-Moufang-Str. 1, D-60438 Frankfurt am Main, Germany}

\date{ }

\begin{abstract}

The decimal logarithm of spontaneous fission half-life of the superheavy
nucleus $^{286}$Fl experimentally determined is $\log_{10} T_f^{exp} (s) =
-0.632$. We present a method to calculate the half-life based on the
cranking inertia and the deformation energy, functions of two independent
surface coordinates, using the best asymmetric two center shell model. 
Spherical shapes are assumed.  In the first stage we study the statics.  At
a given mass asymmetry up to about $\eta=0.5$ the potential barrier has a
two hump shape, but for larger $\eta$ it has only one hump.  The touching
point deformation energy versus mass asymmetry shows the three minima,
produced by shell effects, corresponding to three decay modes: spontaneous
fission, cluster decay, and $\alpha$~decay.  The least action trajectory is
determined in the plane $(R,\eta)$, where $R$ is the separation distance of
the fission fragments and $\eta$ is the mass asymmetry.  We may find a
sequence of several trajectories one of which gives the least action.  The
parametrization with two deformation coordinates $(R,\eta)$ and the radius
of the light fragment, $R_2$, exponentially or linearly decreasing with $R$
is compared with the simpler one, in which $R_2$~=constant and with a
linearly decreasing or linearly increasing $R_2$.  The latter is closer to
the reality and reminds us about the $\alpha $ or cluster preformation at the
nuclear surface.

\end{abstract}


\maketitle

\section{Introduction}
\label{sec:1}

Superheavy (SH) nuclei, with atomic numbers $Z=104-118$, are decaying mainly
by $\alpha $~decay and spontaneous fission.  They have been produced in cold
fusion or hot fusion ($^{48}$Ca projectile) reactions
\cite{khu14prl,ham13arnps,oga11ra,hof11ra,nag11ra,sob11ra,due10prl,mor07jpsjb,oga07jpg,hof00rmp}. 
In a systematic study of $\alpha $-decay energies and half-lives of superheavy
nuclei it was shown \cite{wan15prc} that our semFIS (semiempirical formula
based on fission theory) and UNIV (universal curve) are the best among 18
calculations methods of $\alpha $~decay half-lives.  For some isotopes of
even heavier SHs, with $Z>121$, there is a good chance for cluster decay
modes to compete \cite{p315prc12,p309prl11}.

There are many sources of experimental values for half-lives, $T_f$, of SHs
against spontaneous fission, e.g.,  \cite{aud12cpc1}. 
Among them we found $\log_{10} T_f^{exp} (s) = -3.086, -0.980$ for
$^{282,284}$Cn and $-0.632$ for $^{286}$Fl.  Calculations have been also
performed with different models
\cite{sta13prc,war11prc,smo97pr,smo95pr,bao15jpg,bao13np,san10npa,xu08prc}.

Fission dynamics with Werner-Wheeler nuclear inertia tensor \cite{p326jpg13}
is not leading closer to experiment due to a too small value of inertia; we
tried to improve the agreement between theory and experiment for $^{284}$Cn
by using different laws of variation of mass parameter with fragment
separation distance.  Better results are obtained for $^{282}$Cn with
cranking inertia \cite{p333jpg14} by assuming the most effective split to be
$^{282}Cn \rightarrow ^{130}Pd + ^{152}Dy$.  

In the present work we continue to use the cranking inertia
\cite{bra72rmp,sch86zp,p261epja05} introduced by Inglis \cite{ing54pr}. 
This time we try to find out the least action trajectory in the plane of two
independent variables $(R,\eta)$, where $R$ is the separation distance of
the fragments and $\eta =(A_1 - A_2)/A$ is the mass asymmetry with $A, A_1,
A_2$ the mass numbers of the parent and nuclear fragments.  We assume $A_1
\geq A_2$ hence $\eta \geq 0$.  Consequently both potential energy surfaces
and contour plots (figures like Figs.~\ref{pesflc}, \ref{cflc},
\ref{blg}), function of $(R,\eta)$, will not have the mirror part
corresponding to $A_1 < A_2$.

There are two main terms in the action integral allowing to calculate the
half-life: the total deformation energy and the cranking inertia, both
functions of $(R,\eta)$.  We are using the macroscopic-microscopic method
\cite{str67np} to estimate the deformation energy, expressed as a sum of
Yukawa-plus-exponential (Y+EM) \cite{kra79pr} phenomenological energy,
$E_{Y+E}$, and the shell plus pairing corrections, $\delta E = \delta U +
\delta P$ based on the asymmetric two center shell model (ATCSM)
\cite{ghe03prc,gre96bt}:
\begin{equation} 
E_{def} = E_{Y+E} + \delta E
\end{equation} 
We shall briefly outline the model and discuss the obtained results.

\section{Model}

\subsection{Surface parametrization. Two deformation parameters}

By choosing four independent deformation parameters $R, b_2, \chi _1,\chi
_2$ \cite{pg264pr05} during the deformation from one parent nucleus to two
fission fragments, the surface equation in cylindrical coordinates $\rho ,
z$ is given by 
\begin{equation}
\rho _s^2(z;b_1,\chi _1,b_2,\chi _2)= \left \{ \begin{array}{ccc}
b_1^2-\chi _1 ^2z^2 & , -a_1 < z < z_c \\
b_2 ^2 -\chi _2 ^2 (z-R)^2 & , z_c<z<R+a_2
\end{array} \right.
\end{equation}
where $z_c$ is the position of the crossing plane.

The semiaxes ratio of spheroidally deformed fragments are denoted by $\chi
_1= b_1/a_1$, $\chi _2=b_2/a_2$.  The scalar, $B(R)$, is determined by the
components of the nuclear inertia tensor and the derivatives with respect to
$R$:
\begin{eqnarray} 
B(R) & = & B_{b_2b_2}\left( \frac{db_2}{dR} \right)^2+
2B_{b_2 \chi _1}\frac{db_2}{dR}\frac{d\chi _1}{dR} +2B_{b_2 \chi _2}
\frac{db _2}{dR} \frac{d\chi _2}{dR}+ \nonumber \\ & & 2B_{b _2 R}\frac{db
_2}{dR} + B_{\chi _1 \chi _1}\left(\frac{d\chi _1}{dR}\right)^2+ 2B_{\chi _1
\chi _2}\frac{d\chi _1}{dR}\frac{d\chi _2}{dR}+  \nonumber \\ 
& & 2B_{\chi _1 R}\frac{d\chi _1}{dR}+B_{\chi _2 \chi _2} \left(\frac{d\chi
_2}{dR}\right)^2+ 2B_{\chi _2 R}\frac{d\chi _2}{dR}+ B_{RR} 
\end{eqnarray}

When the two fragments are spheres, $b_2=R_2$, $\chi _1,\chi_2 =1$, meaning
that $\frac{d\chi _1}{dR}=\frac{d\chi _2}{dR}=0$ and the above equation
becomes
\begin{equation} 
B(R) =  B_{b_2b_2}\left( \frac{db_2}{dR} \right)^2+ 
2B_{b _2 R}\frac{db_2}{dR} + B_{RR}  = B_{22} + B_{21} + B_{11} 
\label{br}
\end{equation}
The derivative $\frac{db_2}{dR}=\frac{dR_2}{dR}$ depends only on
geometry. It is a negative quantity since $R_2$ decreases exponentially with
$R$; its absolute values are rather small.

For a given mass asymmetry the final value of the radius of the light
fragment $R_{2f}=r_0A_2^{1/3}$ is well determined.  We assume an exponential
law for the variation with $R$: 
\begin{equation} 
R_2 = R_{2f} + (R_{20} - R_{2f})e^{-k_2\frac{R-R_i}{R_t-R_i}} 
\end{equation}
where $R_{20}=R_0=r_0A^{1/3}$ is equal to the radius of the parent, and the
initial and touching point separation distances are $R_i=R_0 - R_{2f}$ and
$R_t=R_{1f} + R_{2f}$. The radius constant in Y+EM is $r_0=1.16$~fm and
$k_2=4.$ We use this particular value in order to obtain $R_2(x)$ for
$x=(R-R_i)/(R_t-R_i)=1$ very close to the final value $R_2 = R_{2f}$.  When
$k_2=4$, we get $R_2(1)=1.018R_{2f}$, meaning an accuracy of 1.8 ~\%. 
An even larger value of $k_2$ would increase the accuracy but it will also
increase the nuclear inertia, because the shape variation will be faster.
Nuclear inertia is already too large, hence we would not like to increase it  
further.

Previously we took $R_2=R_{2f}$ and consequently we had only one deformation
parameter, $R$, hence $B(R)=B_{RR}(R)$.

We would also like to try two other possibilities: 

(1) Linearly decreasing law from $R_{20}=R_0$ to $R_{2f}=R_e$:
\begin{equation}
R_2 = R_{2f} + (R_{20} - R_{2f})\frac{R_t-R}{R_t-R_i} 
\end{equation} 

(2)  Linearly increasing law from 0 to $R_{2f}=R_e$:
\begin{equation}
R_2 = R_{2f}\frac{R-R_i}{R_t-R_i}
\end{equation} 

For any $R_2$ and  $R_1$, the matching condition at the intersection plane of
the two spheres, gives the solution
\begin{equation} 
z_c = (R_1^2 - R_2^2 + R^2)/(2R)
\end{equation}
where $z_c$ is the distance of the intersection plane from the center of the
heavy fragment.

\subsection{Macroscopic Y+EM energy}

For binary fragmentation with different charge densities, $\rho_{1e}$  and
$\rho_{2e}$ \cite{p80cpc80}, of the Y+EM deformation energy we gave the
details of calculations in Refs.  \cite{p266pr06,p302bb10} :
\begin{eqnarray}
E_{Y+EM} & = (E_Y - E_Y^0)+(E_c - E_c^0) \nonumber \\  & = E_Y^0[B_Y - 1 +
2X(B_c -1)]
\end{eqnarray}
where $E_Y^0=a_2A^{2/3} \{ 1-3x^2+(1+1/x)[2+3x(1+x)]\exp (-2/x) \}$, $E_c^0
= a_cZ^2A^{-1/3}$ are energies corresponding to spherical shape and
$a_2=a_s(1-\kappa I^2)$, $I=(N-Z)/A$, $x=a/R_0$, $R_0=r_0A^{1/3}$.  The
parameters $a_s , \kappa , a_c=3e^2/(5r_0)$, and $r_0$ are taken from 
M\"oller {\it et al.}~\cite{mol95adnd} :
\begin{equation}
B_Y =\frac{E_Y}{E_Y ^0} = \frac{a_{21}}{a_{20}} B_{Y1} +
\frac{\sqrt {a_{21} a_{22}}}{a_{20}} B_{Y12} +
\frac{a_{22}}{a_{20}} B_{Y2}
\end{equation}
The relative Yukawa and Coulomb energies $B_Y=E_Y/E_Y^0$, $B_c=E_c/E_c^0$
are functions of the nuclear shape; with axially-symmetric shapes they 
are expressed by triple integrals.
In a similar way the Coulomb relative energy is given by
\begin{equation}
B_c =\frac{E_c}{E_c ^0} = \left ( \frac{\rho _{1e}}{\rho _{0e}}
\right ) ^2 B_{c1} + \frac{\rho _{1e} \rho _{2e}}{\rho _{0e} ^2}
B_{c12} + \left ( \frac{\rho _{2e}}{\rho _{0e}} \right ) ^2 B_{c2} 
\end{equation}
where again one can see the self-energies $B_{c1}, B_{c2}$ and the
interaction $B_{c12}$.

\subsection{Shell and pairing corrections}

The input is obtained from the ATCSM \cite{ghe03prc}; at every pair of
coordinates $(R,\eta)$ we get a sequence of doubly
degenerate discrete energy levels $\epsilon _i = E_i /\hbar \omega^0_0$ in
units of $\hbar \omega^0_0 = 41 A^{-1/3}$, arranged in order of increasing
energy. In units of $\hbar \omega_0^0$ the shell corrections are 
determined as
\begin{equation}
\delta u(n, R,\eta) = \sum_{i=1}^n 2\epsilon_i(R,\eta) -
\tilde{u}(n, R,\eta)
\end{equation}
with $n=N_p/2$ particles and $\tilde{u}$ the total energy of the uniform
level distribution calculated with Strutinsky's \cite{str67np} procedure. 
Then we add the contributions from protons and neutrons 
$\delta u = \delta u_p + \delta u_n$.

For pairing corrections we have first to solve the BCS \cite{bar57pr}
system of two equations with two unknowns, Fermi energy $\lambda$ and the
pairing gap $\Delta$, 
\begin{equation} 0 = \sum_{k_i}^{k_f}\frac{\epsilon_k
-\lambda}{\sqrt{(\epsilon_k
-\lambda)^2+\Delta^2}} 
\end{equation} 
\begin{equation}
 \frac{2}{G} = \sum_{k_i}^{k_f}\frac{1}{\sqrt{(\epsilon _k
-\lambda)^2+\Delta^2}} \end{equation} where $k_i=Z/2-n+1, \; \; k_f=Z/2+n'$
for proton levels, and \begin{equation} \frac{2}{G} \simeq
2\tilde{g}(\tilde{\lambda})\ln \left (\frac {2\Omega}{\tilde{\Delta}} 
\right) 
\end{equation}
assuming that for protons $Z/2$ levels are occupied with $n$ levels below
and $n'$ above Fermi energy contributing to pairing, $n=n'=\Omega
\tilde{g_s} /2$.  The cutoff energy, $\Omega \simeq 1 \gg
\tilde{\Delta}=12/\sqrt{A}\hbar\omega_0^0$.

Occupation probability by a quasiparticle ($u_k^2$) or hole ($v_k^2$) is
given by
\begin{equation}
v_k^2 = \left
[1-(\epsilon_k -\lambda)/E_k \right ]/2 ; \; \; u_k^2 = 1-v_k^2
\end{equation}
The quasiparticle energy is expressed as
\begin{equation}
E_\nu=\sqrt{(\epsilon _\nu -\lambda)^2+\Delta^2}. 
\end{equation}
The pairing correction, $\delta p=p-\tilde{p}$,
represents the difference between the pairing correlation
energies for the discrete level distribution
\begin{equation}
p = \sum_{k=k_i}^{k_f}2v_k^2\epsilon_k -2\sum_{k=k_i}^{Z/2}\epsilon_k
- \frac{\Delta^2}{G}
\end{equation}
and for the continuous level distribution
\begin{equation}
\tilde{p} = -(\tilde{g}\tilde{\Delta^2})/2 =-(\tilde{g}_s\tilde{\Delta
^2})/4
\end{equation}
Compared to shell correction, the pairing correction is out of phase
and smaller in amplitude, leading for $\eta=$~constant
to a smoother total curve $\delta e (R) =
\delta u (R) + \delta p (R) $, where $\delta p = \delta p_p + \delta p_n$. 

\begin{figure}[ht]
\begin{center}
\includegraphics[width=12cm]{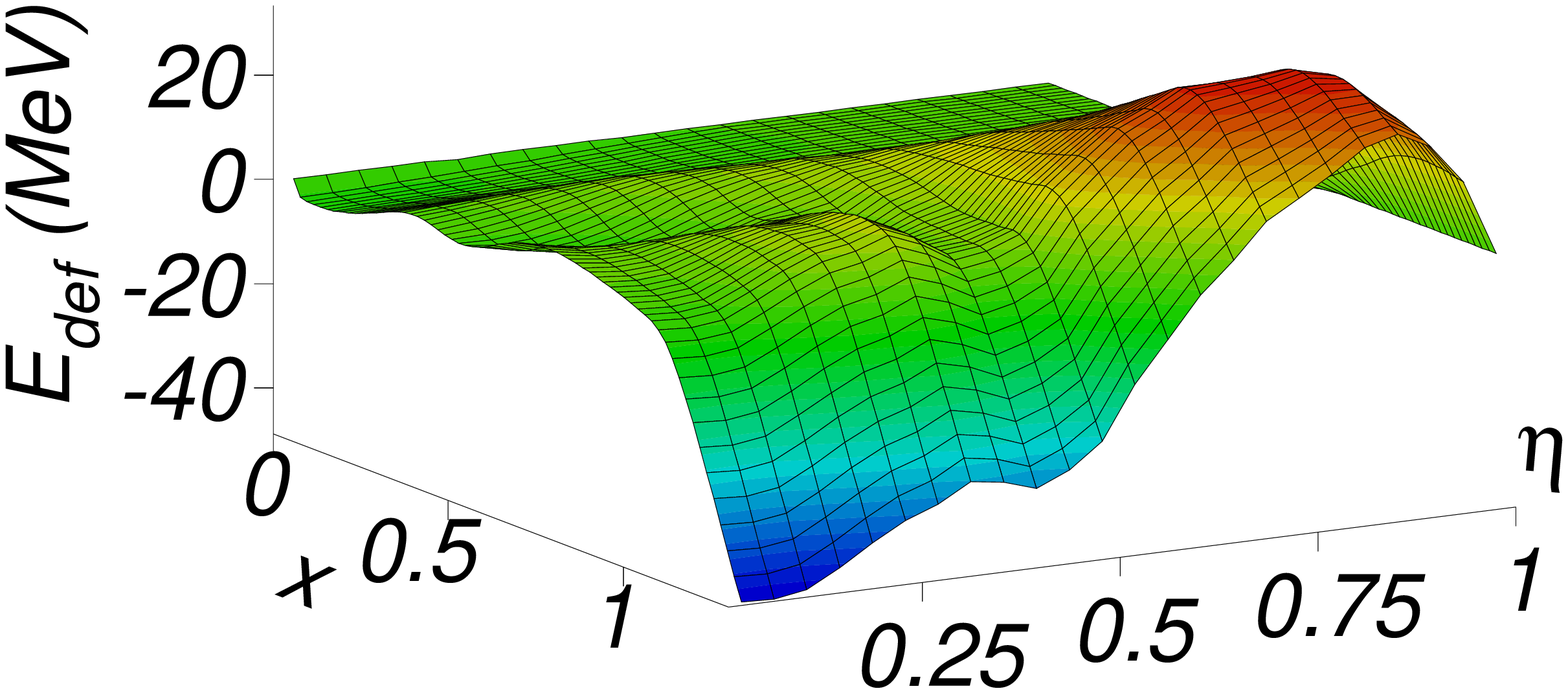} \vspace*{-0.9cm} 

\includegraphics[width=12cm]{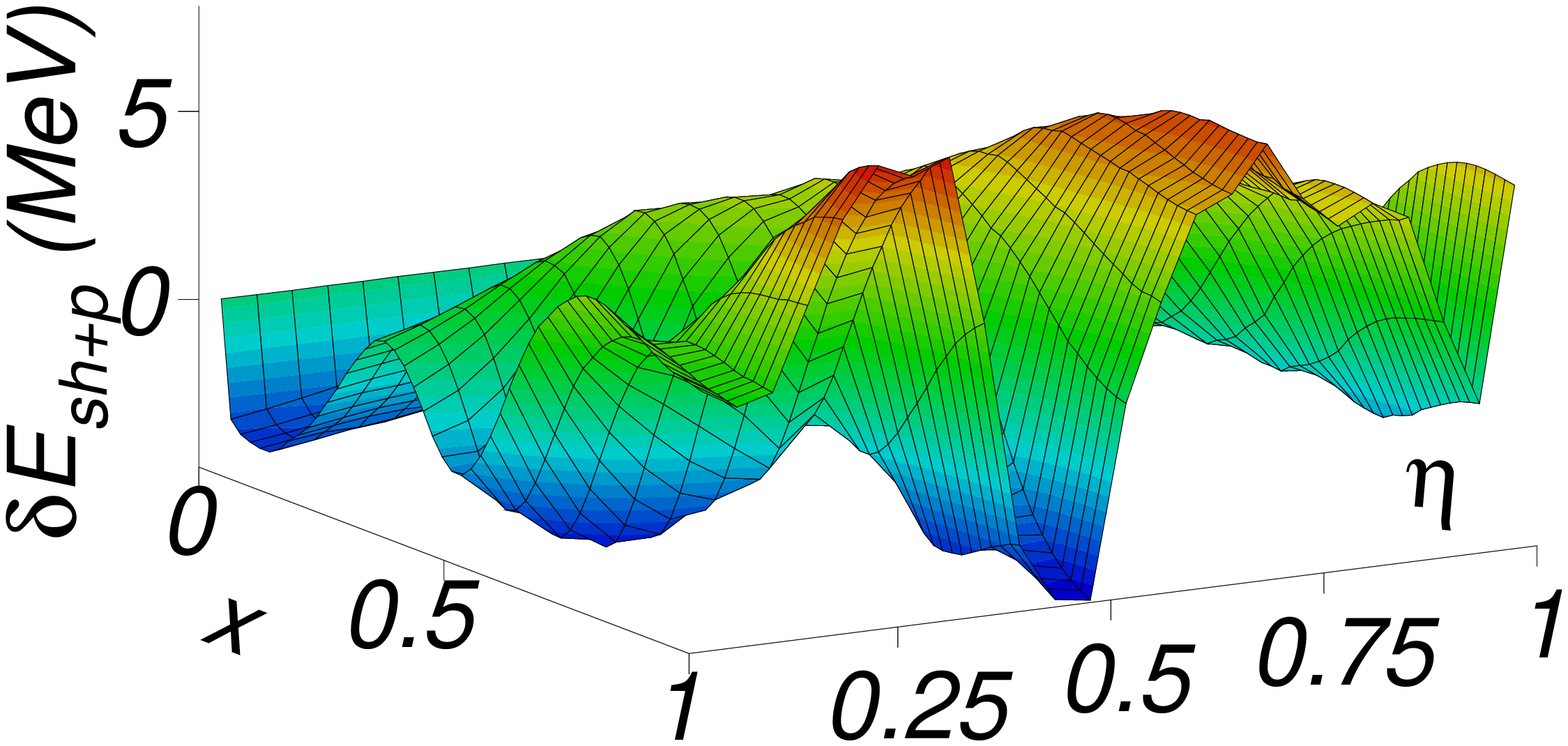} \vspace*{-0.9cm} 

\includegraphics[width=12cm]{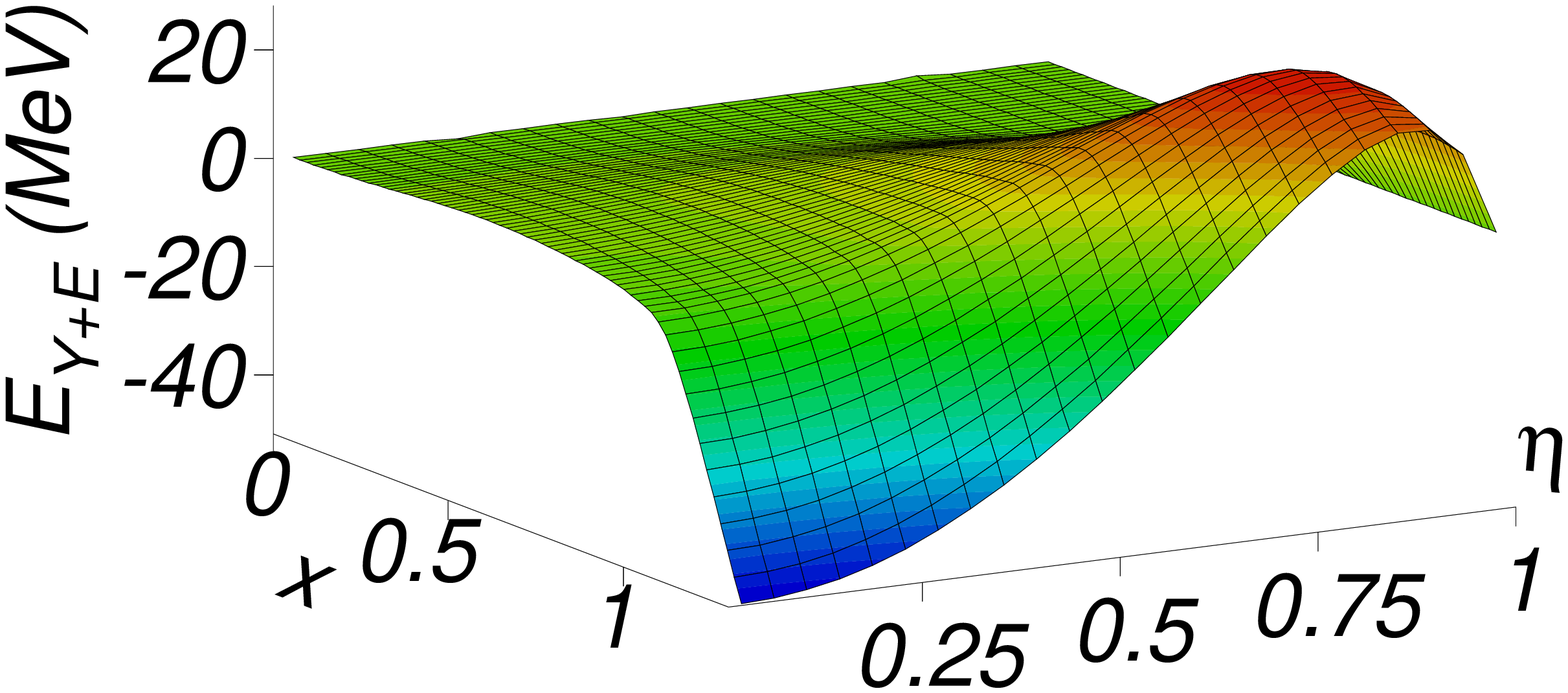}  
\end{center}
\caption{(Color online) PES of $^{286}$Fl vs $(R-R_i)/(R_t - R_i) \geq 0$
and $\eta = (A_1-A_2)/(A_1+A_2)$.  Y+EM (bottom), Shell + Pairing
corrections (center), and total deformation energy (top).
\label{pesflc}} 
\end{figure}

\begin{figure}[ht]
\centerline{\includegraphics[width=12cm]{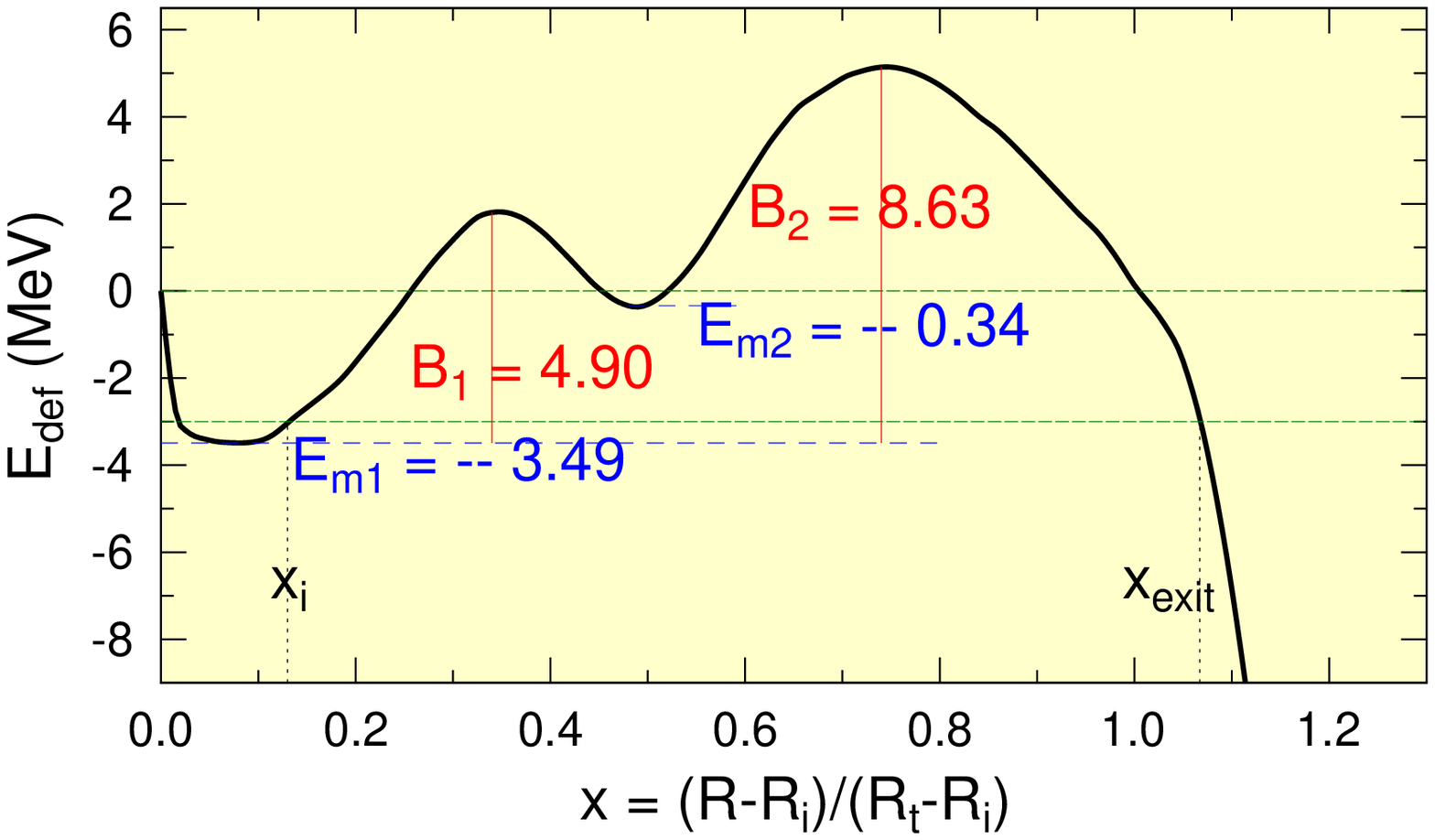}} 
\caption{(Color online) Deformation energy of $^{286}$Fl symmetrical
fission.  Important characteristics of the two humped barrier: first and
second minima, $E_{m1}, E_{m2}$, first and second barrier height, $B_1,
B_2$, and the two turning points, $x_i, x_{exit}$. The difference in energy
from the exit dashed line and the deepest minimum, $E_{m1}$, is the
zero-point vibration energy, $E_v$.
\label{fl0}} 
\end{figure}

\subsection{Total deformation energy}

\begin{figure}
\centerline{\includegraphics[width=11cm]{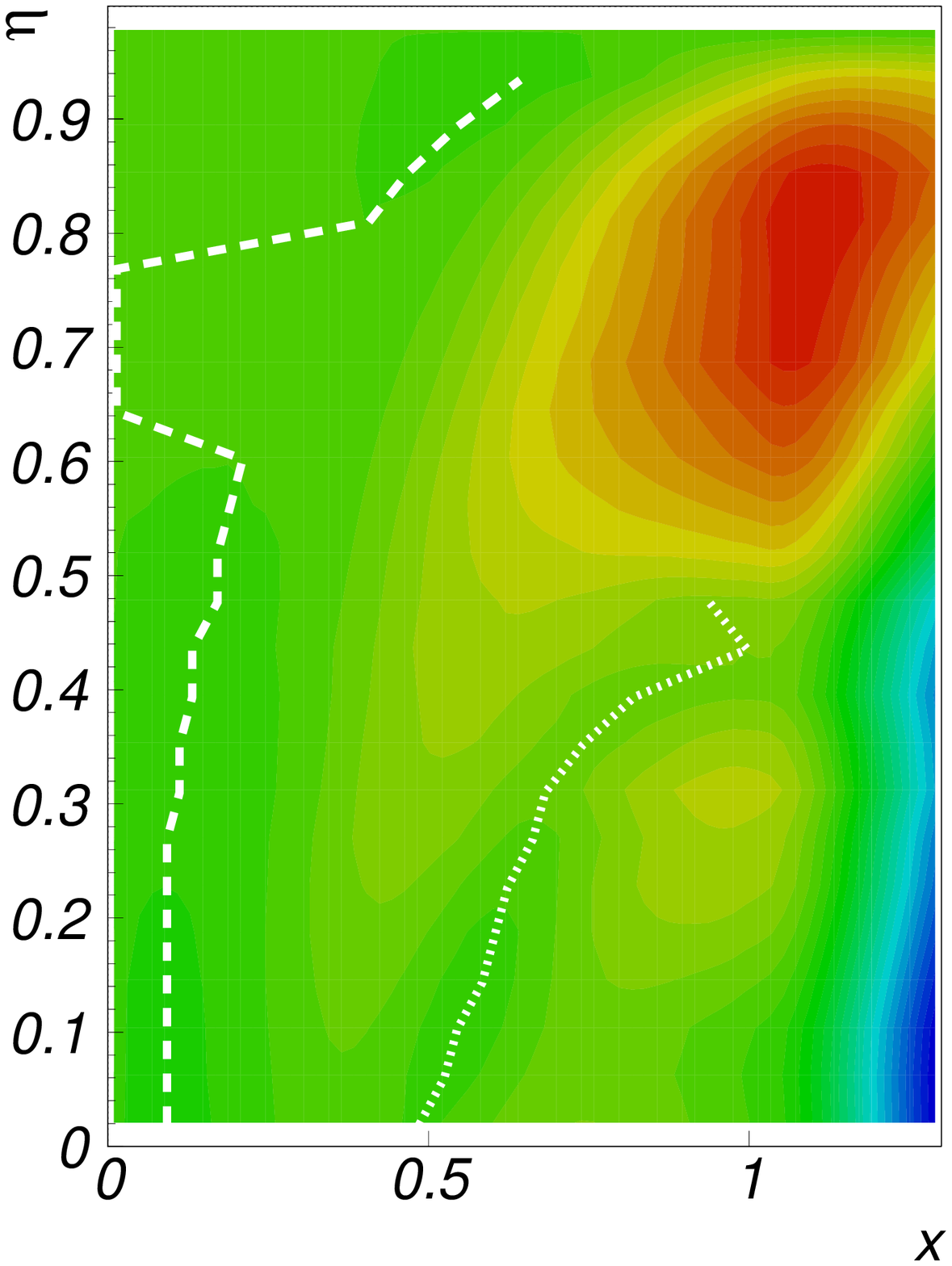}} 
\vspace*{-1cm} 
\caption{(Color online) Contour plot of deformation energy of $^{286}$Fl
shown as a PES in the upper panel of Fig.~\ref{pesflc}.  The first and
second minima of deformation energy at every value of mass asymmetry are
plotted with dashed and dotted white lines.
\label{cflc}} 
\end{figure}

After subtracting the values of deformation energy of the parent we can make
the final sum
\begin{equation}
E_{def}=E_{Y+E}+\delta E_{sh+p}
\end{equation}
Potential energy surfaces (PES) and contour plots for spontaneous fission of
$^{286}$Fl are shown in figures~\ref{pesflc} and~\ref{cflc}.  In figure
~\ref{cflc} we also show with white dashed and dotted lines the minima of
deformation energy at every mass asymmetry (see also the Table~\ref{tab}). 
A cut in PES at symmetry, $\eta=0$, is plotted in Fig.~\ref{fl0}, where
one can see not only the total energy but also the important characteristics
given in Table~\ref{tab}: first and second minima ($E_{m1},E_{m2}$), first
and second barrier height ($B_1, B_2$), and the two turning points $x_i,
x_{exit}$, taking care to allow for a small value of zero-point vibration
energy, $E_v$, from the deepest minimum $E_{m1}$ to the exit line.  Two deep
minima in the shell plus pairing correction energy correspond to the doubly
magic fragments $^{132}$Sn (near symmetry) and $^{208}$Pb (at a value of
$\eta$ about 0.5) which are responsible for spontaneous fission and cluster
decay, respectively.
\begin{table}[hbt] 
\caption{Statics. Minima and maxima of deformation energy in MeV 
for fission of $^{286}$Fl. x$_{exit}$ corresponds to $E_v=0.$
\label{tab}} 
\begin{center}
\begin{ruledtabular}
\begin{tabular}{c|cc|cc|cc|cc|c}
$\eta$ &x &1st min. &x &1st max. & x &2nd min. & x& 2nd max. &x$_{exit}$\\
\hline
0.000 & 0.074 & -3.490 &0.352 &1.810 & 0.482 &-0.340 &0.741 & 5.143 & 1.013 \\
0.043 & 0.074 & -3.347 &0.352 &2.260 & 0.519 &-1.027 &0.760 & 4.079 & 0.990 \\
0.087 & 0.074 & -3.190 &0.371 &2.843 & 0.556 &-1.431 &0.779 & 4.031 & 1.024 \\
0.130 & 0.074 & -3.025 &0.390 &3.573 & 0.576 &-1.382 &0.836 & 5.398 & 1.085 \\
0.174 & 0.074 & -2.826 &0.409 &4.484 & 0.595 &-0.753 &0.893 & 7.365 & 1.118 \\
0.217 & 0.074 & -2.600 &0.429 &5.518 & 0.634 & 0.434 &0.950 & 9.282 & 1.143 \\
0.261 & 0.075 & -2.327 &0.467 &6.486 & 0.654 & 1.707 &0.971 & 9.896 & 1.154 \\
0.304 & 0.094 & -2.091 &0.506 &7.200 & 0.674 & 3.790 &0.980 &11.674 & 1.176 \\
0.348 & 0.113 & -1.754 &0.526 &7.941 & 0.733 & 4.256 &0.996 & 8.770 & 1.163 \\
0.391 & 0.113 & -1.408 &0.566 &8.526 & 0.831 & 3.323 &0.982 & 4.700 & 1.134 \\
0.435 & 0.133 & -0.988 &0.626 &9.825 & & & & & 1.150 \\
0.478 & 0.152 & -0.720 &0.648 &10.448& & & & & 1.182 \\
0.522 & 0.173 & -0.424 &1.036 &15.086& & & & & 1.263 \\
0.565 & 0.174 & -0.186 &1.044 &20.224& & & & & 1.301 \\
0.609 & 0.195 & -0.023 &1.052 &25.017& & & & & 1.404 \\
0.652 & 0.000 &  0.000 &1.062 &27.948& & & & & 1.479 \\
0.696 & 0.000 &  0.000 &1.074 &31.447& & & & & 1.588 \\
0.739 & 0.000 &  0.000 &1.088 &31.786& & & & & 1.689 \\
0.783 & 0.000 &  0.000 &1.084 &32.317& & & & & 1.845 \\
0.826 & 0.396 & -0.047 &1.104 &33.336& & & & & 2.125 \\
0.870 & 0.469 & -0.284 &1.131 &31.884& & & & & 2.550 \\
0.913 & 0.529 & -0.630 &1.146 &25.720& & & & & 3.130 \\
0.956 & 0.649 & -1.080 &1.182 &15.717& & & & & 4.470 \\
\end{tabular}
\end{ruledtabular}
\end{center}
\end{table}
If we use in graphics $x=(R-R_i)/(R_t-R_i)$ instead of $R$ 
then for $^{286}$Fl the interval of variation will be $x=(0, 1)$.
For the initial parent nucleus one may have either $x=0$ or/and $\eta =1$.
This is the reason why the dashed line ends up at the value of $\eta
=0.956$. In present calculations we have used 66 values of $x$ from $0$ to
$1.3$ and 24 values of $\eta$ from $0$ to $1$.

For mass asymmetry  $\eta \leq 0.435$ we obtain a double hump potential
barrier as shown in 
Table~\ref{tab}, where the position of minima and maxima as well as the
height of the two barriers (local maximum minus the ground state minimum)
and the second minimum are also given.  
The deepest minimum, which should be taken as the ground state corresponds
to $x=0.074 \ \eta=0.00$, where $E_{def}=-3.49$~MeV.
Assuming zero point vibration energy $E_v=0$, the exit point from the
barrier is also given.  Initially, at $\eta=0$, the exit point is about
x$_{exit} =0.990 $ (see Fig.~\ref{fl0}).  The existence of a two hump
barrier for $\eta \leq 0.435$ is mainly related to the importance of the two
double magic fragments $^{132}$Sn and $^{208}$Pb.  The limit observed from
Table~\ref{tab} is not far from $\eta = (208-78)/286=0.4545$.  Both from the
Fig.~\ref{cflc} and the Table~\ref{tab} we can see that at a given mass
asymmetry up to about $\eta=0.5$ the potential barrier has a two hump shape,
but for larger $\eta$ it has only one hump.  This fact is related to the
presence of Businaro-Gallone mountain \cite{bus55nc} as well as to the level
densities at a large value of $\eta $ and $x=(R-R_i)/(R_t - R_i)$.
\begin{figure}
\centerline{\includegraphics[width=10cm]{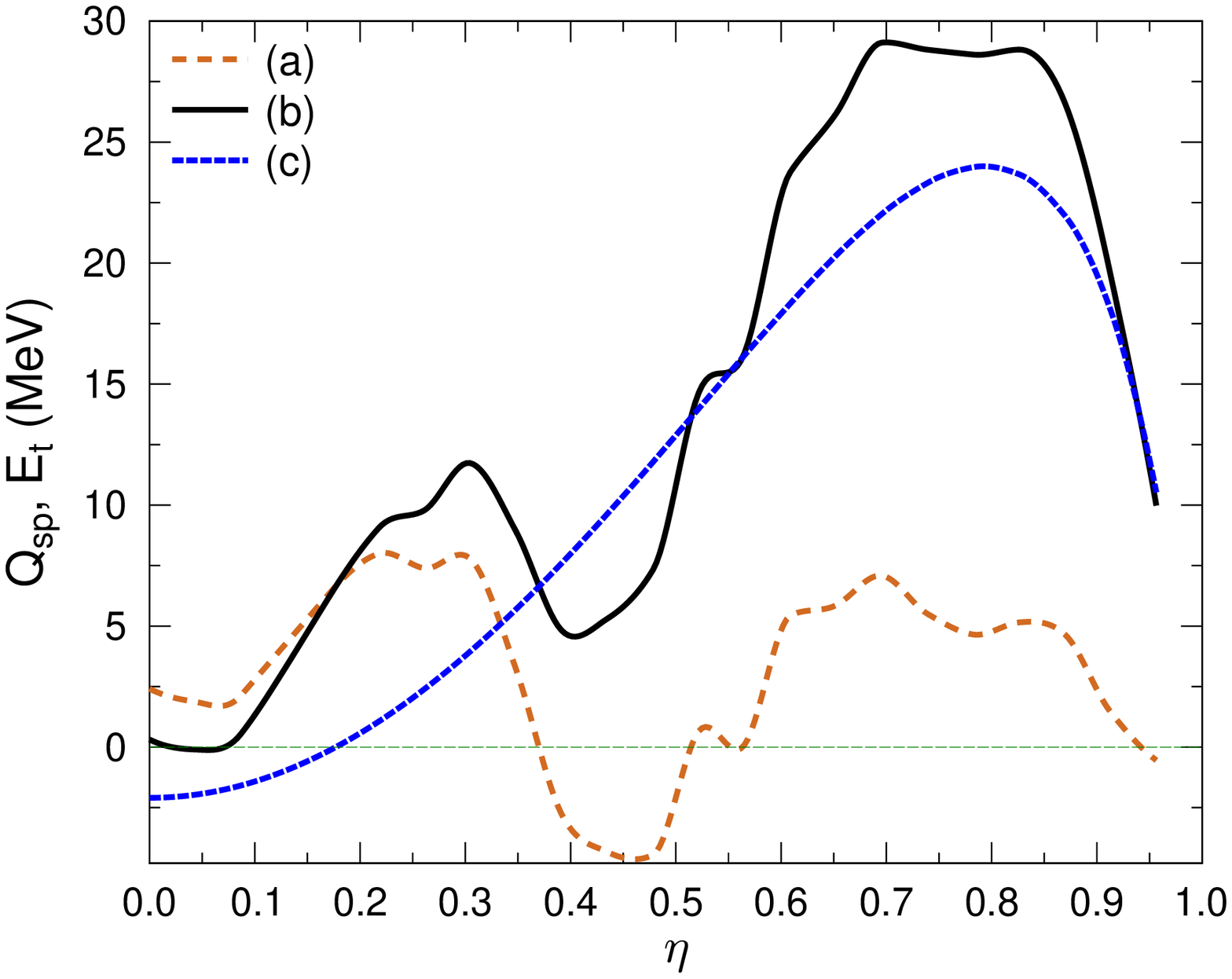}} 
\caption{(Color online) The touching point deformation energy, $E_t$ (b), its
macroscopic part, $E_{Y+E}=E_t - Q_{sp}$ (c), and the contribution of shell and
pairing corrections, $Q_{sp}$ (a), for SF of $^{286}$Fl, versus mass asymmetry.
\label{cft}} 
\end{figure} 
The macroscopic part (Y+EM) of deformation energy (heavy dashed blue line)
with a maximum at $\eta =0.826$, and the total value, $E_t$, including the
contribution of shell and pairing corrections, $Q_{sh}$, for SF of
$^{286}$Fl, versus mass asymmetry is shown in Fig.~\ref{cft}.  Around the
mass symmetry, $\eta=0.0$, up to $\eta=0.177$ we have $E_{Y+E} < 0.0$.  From
the minima of $Q_{sp}$ (a), we can see the three main regions in the order of
increasing value of $\eta$ around the doubly magic daughters $^{132}$Sn and
$^{208}$Pb as well as the doubly magic emitted $^4$He.  The corresponding
valleys on the PES are favorable to spontaneous fission, cluster decay, and
$\alpha$~decay.

\subsection{Cranking inertia} \begin{figure}
\centerline{\includegraphics[width=10cm]{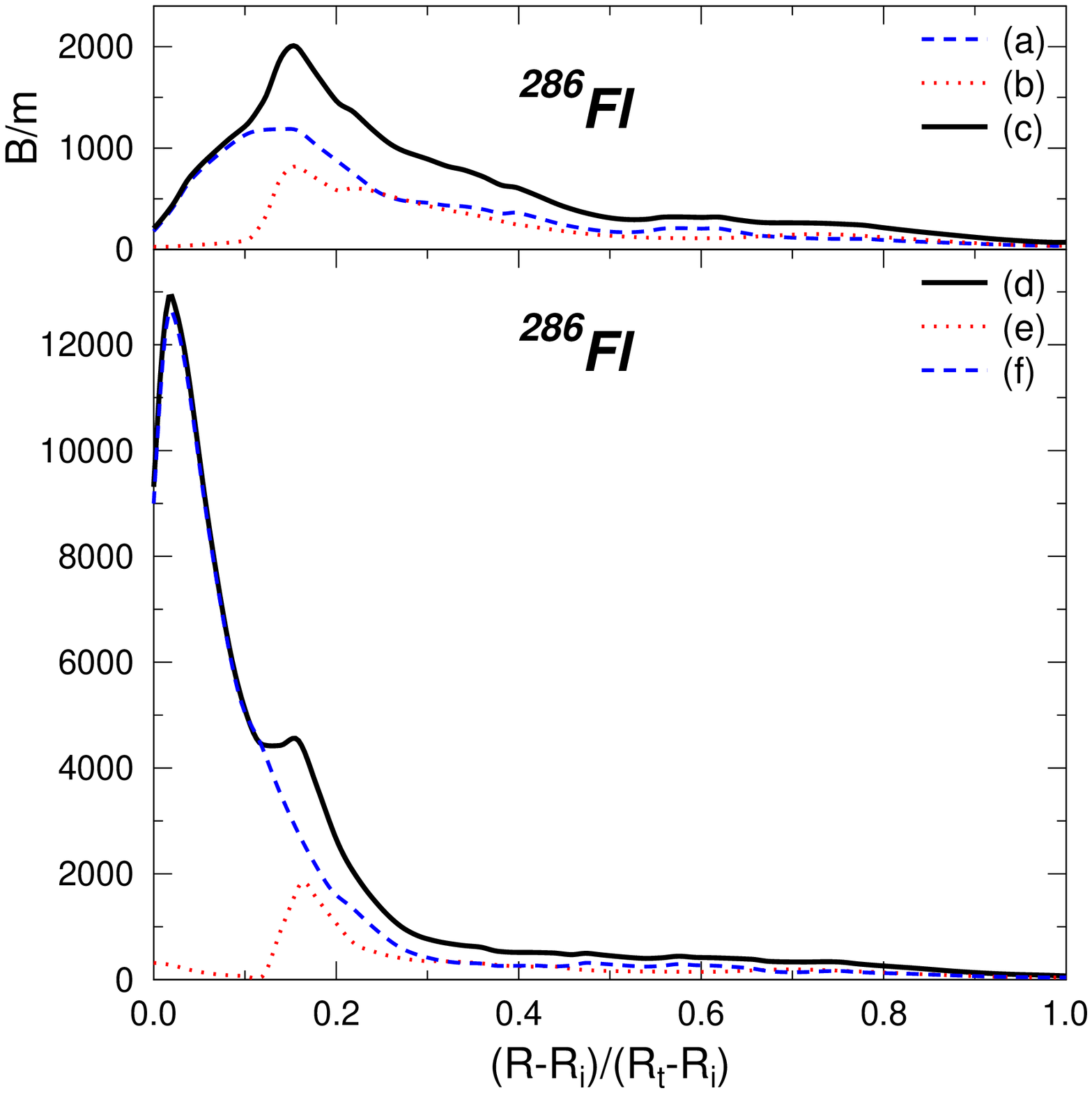}} 
\caption{(Color online) Cranking inertia with proton and neutron
contributions.  Top: fission of $^{286}$Fl with $^{132}$Sn light fragment;
$R_2=$~constant.  (a) neutrons contributions; (b) protons contributions; (c)
total.  Bottom: symmetrical spontaneous fission of $^{286}$Fl; exponential
decrease of $R_2$.  (d) total; (e) protons contributions; (f) neutrons
contributions.  $(R-R_i)/(R_t-R_i)$ and B/m are dimensionless quantities. 
\label{crm}} 
\end{figure} 
\begin{figure}
\centerline{\includegraphics[width=10cm]{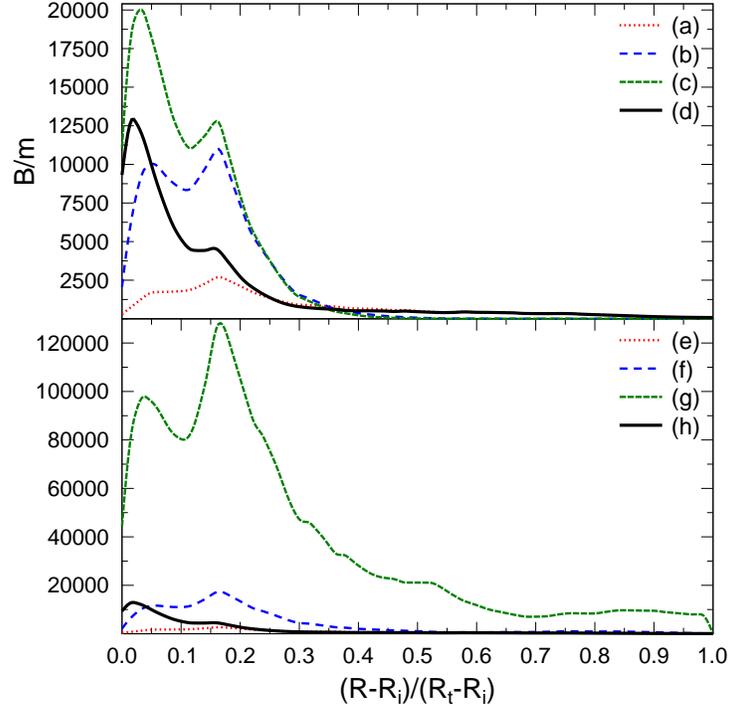}} 
\caption{(Color online) Cranking inertia components for symmetrical fission
of $^{286}$Fl.  Two independent deformation coordinates $(R,R_2)$.  $R_2$
decreases exponentially with $R$.  Top.  Three components and the total: (a)
$B_{RR}=B_{11}$; (b) $|B_{12}|$~($B_{12}$ is a negative quantity); (c)
$B_{22}$; (d) $B$.  $B_{11}=B_{RR}, B_{12}=2B_{RR_2}\frac{dR_2}{dR},
B_{22}=B_{R_2R_2}\left(\frac{dR_2}{dR}\right)^2$.  Bottom.  Three components
and the total: (e) $B_{RR}$; (f) $B_{R2R}$; (g) $B_{R2R2}$; (h) $B$. 
$(R-R_i)/(R_t-R_i)$ and B/m are dimensionless quantities.
\label{b3c}}
\end{figure} 
\begin{figure} 
\centerline{\includegraphics[width=12cm]{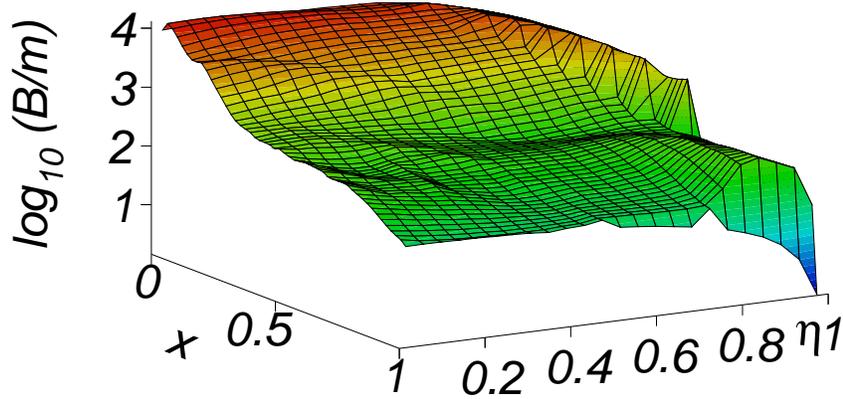}}
\caption{(Color online) Decimal logarithm of nuclear inertia, $\log _{10}
(B/m)$, for fission of $^{286}$Fl.  B/m, x and $\eta$ are dimensionless
quantities.  
\label{blg}} 
\end{figure}

\begin{figure}
\centerline{\includegraphics[width=11cm]{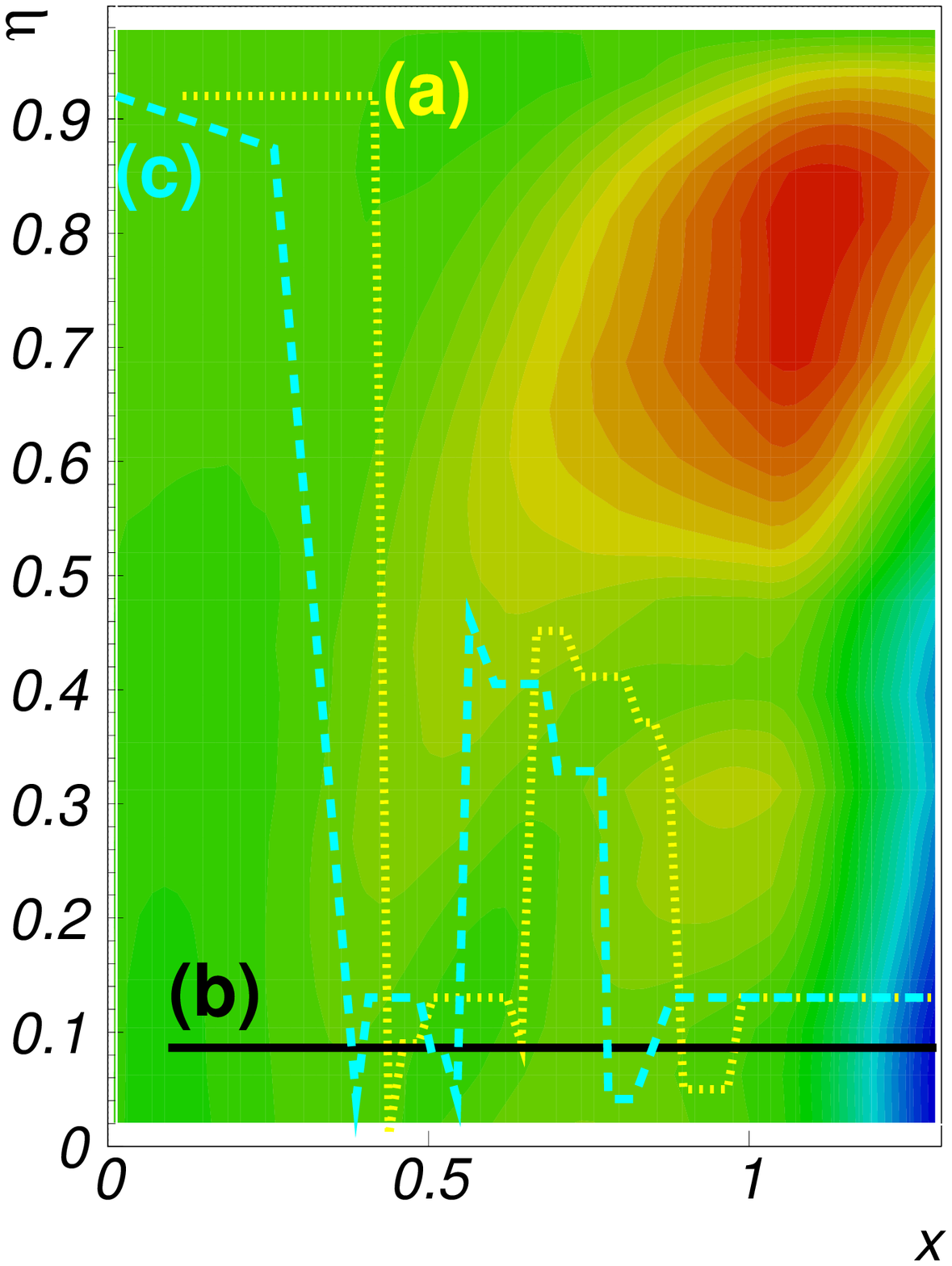}} 
\vspace*{-2cm}
\caption{(Color online) Three least action trajectories on the contour plot
of deformation energy of $^{286}$Fl: (a) yellow dotted-line for variable
$R_2$; (b) black solid line for $R_2=$~constant, and (c) cyan dashed-line
for linearly increasing $R_2$.  x and $\eta$ are dimensionless quantities.
\label{path}} 
\end{figure} 

According to the cranking model, after including the BCS pairing
correlations \cite{bar57pr}, the inertia tensor \cite{bra72rmp} is given by
\begin{equation}
B_{ij} =2\hbar^2\sum_{\nu \mu} \frac{\langle \nu|\partial H/\partial
\beta_i|\mu \rangle \langle \mu|\partial H/\partial \beta_j|\nu
\rangle}{(E_\nu +E_\mu)^3}(u_\nu v_\mu +u_\mu v_\nu)^2  
\label{eq3}
\end{equation}
where $H$ is the single-particle Hamiltonian allowing to determine the
energy levels and the wave functions $|\nu \rangle$, $u_\nu^2$, $v_\nu^2$
are the BCS occupation probabilities, $E_\nu$ is the quasiparticle energy,
and $\beta_i, \beta_j$ are the independent shape coordinates.

Again we follow the procedure for proton and neutron levels and the final
result is obtained by adding the two contributions.
As already mentioned above, for two independent shape coordinates we have
\begin{equation}
B(R)=B_{RR}(R,R_2)+
2B_{RR_2}\frac{dR_2}{dR} +
B_{R_2R_2}\left(\frac{dR_2}{dR}\right)^2 = B_{11} + B_{12} + B_{22} 
\end{equation}
where $B_{11}=B_{RR}, B_{12}=2B_{RR_2}\frac{dR_2}{dR},
B_{22}=B_{R_2R_2}\left(\frac{dR_2}{dR}\right)^2$.
In the lower and upper panels of Fig.~\ref{crm} we plotted $B/m$ --- the
cranking inertia in units of the nucleon mass $m$ for symmetrical fission of
$^{286}$Fl and that with the light fragment $^{132}$Sn and $R_2$~constant,
respectively.  One can see that a major contribution comes from the neutrons
[heavy  dashed blue line (a)]. Also, when $R_2$ is decreasing exponentially, 
the inertia is much higher than in the case of $R_2$~constant.  
In Fig.~\ref{b3c}
we compare the three components of nuclear inertia for symmetrical
spontaneous fission of $^{286}$Fl.  The very high value of $B_{R_2R_2}$
[$B_{22}$ with green dashed curve (g) at the bottom] becomes smaller when
multiplied by $\left(\frac{dR_2}{dR}\right)^2$ [green dashed curve (c) at the
top].  On the other hand the value of the component $B_{R_2R}$
[blue dashed curve (f) at the bottom] remains practically at an intermediate
level when multiplied by $\frac{dR_2}{dR}$ leading to (b), i.e., $|B_{12}|$.

For minimization of the least action trajectory in the plane $(R,R_2)$ we
need not only $B_{RR}$ but also the values of $B_{R_2R_2}, B_{R_2R}$ in
every point of a grid of 66$\times$24 for 66 values of $(R-R_i)/(R_t-R_i)$
and 24 values of $\eta = (A_1-A_2)/A$ or $R_{2f}$.

The decimal logarithm of $B/m$ function of $(R,\eta)$ is given in
Fig.~\ref{blg} as a three-dimensional plot.  
At the touching point and beyond, $R \geq
R_t$, one should get the reduced mass: $B(R \geq R_t) = mA_1A_2/A$. 
Generally speaking the values of $B/m$ are higher where the deformation
energy is low.  Consequently we expect a dynamical path (Fig.~\ref{path})
very different from the statical one shown in Fig.~\ref{cflc} with a white
dashed line.

\subsection{Half-life}

The half-life of a parent nucleus $AZ$ against the split into a light
fragment $A_2Z_2$ and a heavy fragment $A_1 Z_1$ is given by
\begin{equation}
T = [(h \ln 2)/(2E_{v})] exp(K_{ov} + K_{s})
\end{equation}
and is calculated by using the Wentzel--Kramers--Brillouin (WKB) 
quasiclassical approximation, according to
which the action integral is expressed as
\begin{equation}
K=\frac{2\sqrt{2m}}{\hbar}\int_{R_a}^{R_b}
\{[(B(R)/m)][E_{def}(R)-E_{def}(R_a)]\}^{1/2}dR
\end{equation}
with $B=$ the cranking inertia, $K=K_{ov}+K_s$, and the $E(R)=E_{def}$ 
potential energy of deformation.  $R_a$ and $R_b$ are the turning points of
the WKB integral where $E_{def} = E_{def}(R_a) = E_{def}(R_b) $.  The two
terms of the action integral $K$, correspond to the overlapping ($K_{ov}$)
and separated ($K_s$) fragments.  We can use the relationship
\begin{equation} 
\log_{10} T = 0.43429(0.4392158S_{ab}) -20.8436 - \log_{10} E_v 
\end{equation} 
where 
\begin{equation} 
S_{ab} = \int _{R_a}^{R_b}
\{[(B(R)/m)][E_{def}(R)-E_{def}(R_a)]\}^{1/2} dR 
\end{equation}

For $^{286}$Fl and $r_0=1.16$~fm (Y+EM) we have $R_0=r_0A^{1/3}=7.6427$~fm,
$R_{1s}=r_0A_{1s}^{1/3}=6.066$~fm, $R_{2s}=r_0A_{2s}^{1/3}=6.066$~fm,
$R_i=R_0-R_{2s}=1.5767$~fm, $R_t=R_{2s}+R_{2s}=12.132$~fm, where the
subscript s stands for symmetry ($\eta=0$). 

\section{Results}
We started to calculate the half-life by choosing for the beginning the
simplest trajectory in the plane $(x, \eta)$, namely $\eta =$~constant.  The
results are shown in Table~\ref{tab2} for four such trajectories.  The
zero-point vibration energy is quite high $4.2835-5.0220$, with a minimum at
$\eta =0.0870$.  From Table~\ref{tab} the corresponding $x_{exit}$
should be smaller than $1.07$.  We continue with least action trajectory, in
which the first guess for the exit point could be not far from this value of
$\eta=0.087$. 

In Fig.~\ref{path} we represent three fission paths, (a), (b), and (c). The
least action trajectory (a) (yellow dotted line) was obtained when the radius of
the light fragment, $R_2$, was exponentially decreased down to the final
value.  In this case, in order to reproduce the experimental value of $T_f$
when using $E_v=0.5$~MeV, it was necessary to diminish substantially the two
components, $B_{12}$ and $B_{22}$ of the cranking inertia tensor.  By taking
$R_2=$~constant (b), as in Table~\ref{tab2}, the dynamical trajectory is
simply a solid straight line.  The best results are obtained when $R_2$ is
linearly increasing leading to the (c) cyan dashed-line and reproducing the
experimental fission half-life with a reasonable zero-point vibration energy
$E_v=0.685$~MeV compared to $E_v=1.361$~MeV for the path (c) with $R_2=$~constant. 

\begin{figure} 
\centerline{\includegraphics[width=14cm]{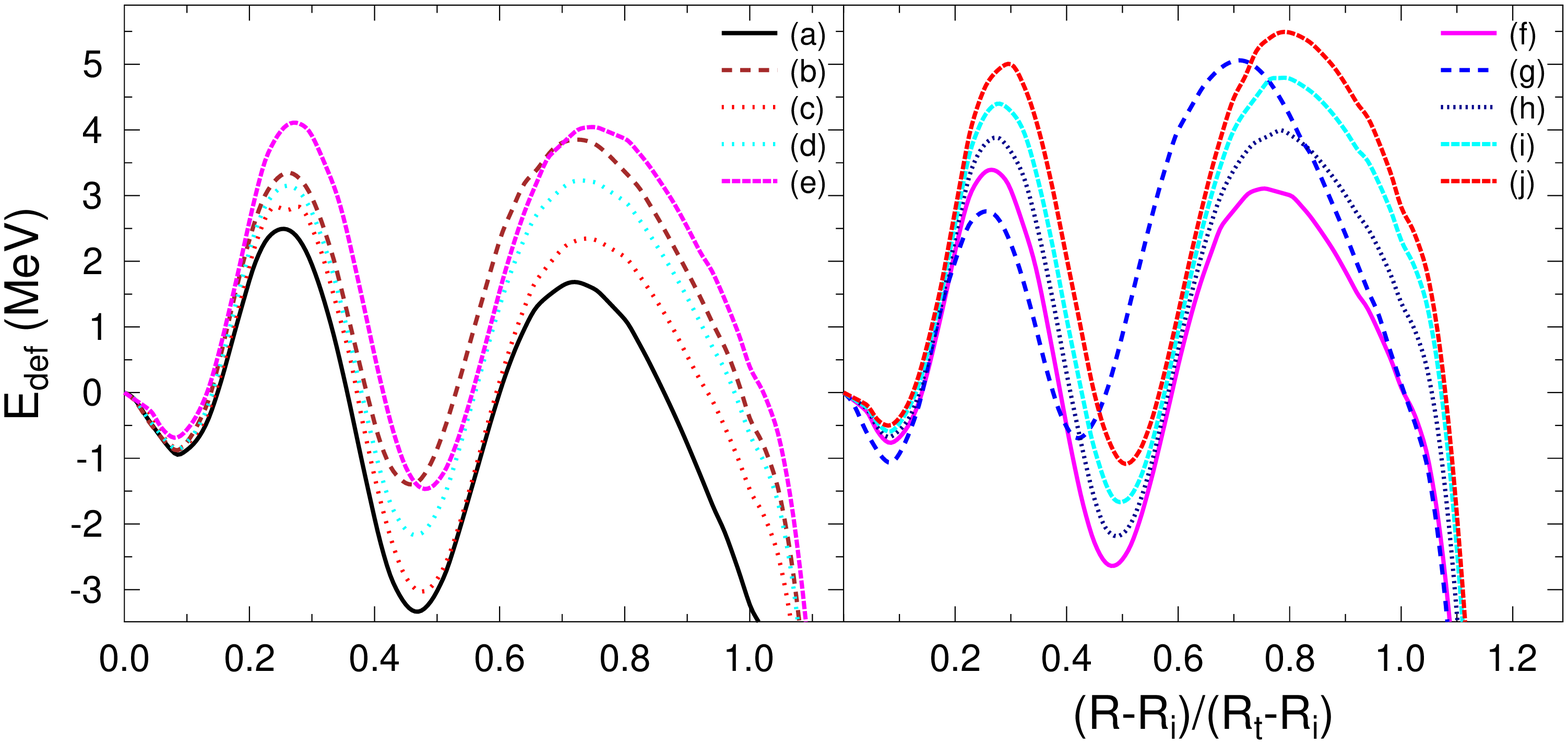}} 
\caption{(Color online) Fission barriers for ten different combinations of
fragments.  The light fragments are the following: (a) $^{134}$Sn; (b)
$^{136}$Xe; (c) $^{132}$Sn; (d) $^{134}$Te; (e) $^{130}$Te; (f) $^{130}$Sn;
(g) $^{143}$La; (h) $^{128}$Sn; (i) $^{126}$Sn; (j) $^{124}$Sn.  Spontaneous
fission of $^{286}$Fl.
\label{10b}} 
\end{figure}
\begin{table}[hbt] 
\caption{Dynamics.  The optimum value of the parameter
zero-point vibration energy, $E_v$, used to reproduce the experimental value
of $^{286}$Fl spontaneous fission half-life, $\log_{10} T_f^{exp} (s)
=-0.63$.  The simplest trajectories, $\eta =$~constant, are used in the plane
$(R,\eta)$.  
\label{tab2}} 
\begin{center} 
\begin{ruledtabular}
\begin{tabular}{ccc} $\eta$ & $E_v$ (MeV) & $\log_{10}T_f (s) $\\ \hline 
0.0000 &5.0220 & -0.63\\ 
0.0430 &4.3909 & -0.63\\ 
0.0870 &4.2835 & -0.63\\ 
0.1304 &4.9450 & -0.63\\ 
\end{tabular}
\end{ruledtabular} 
\end{center} 
\end{table} 
\begin{table}[hbt]
\caption{Dynamics.  The optimum value of the parameter zero-point vibration
energy, $E_v$, used to reproduce the experimental value of $^{286}$Fl
spontaneous fission half-life, for a given split, using a shape
parametrization with $R_2=$~constant.  \label{tab3}} 
\begin{center}
\begin{ruledtabular} 
\begin{tabular}{ccccc} 
$\eta$ & $A_2$ & $Z_2$ & $E_v$ (MeV) \\ \hline 
0.0769 &132 & 50 & 1.3612\\ 
0.0909 &130 & 52 & 1.4278\\
0.0629 &134 & 50 & 1.4762\\ 
0.0629 &134 & 52 & 1.5690\\ 
0.0490 &136 & 54 & 2.0916\\ 
\end{tabular} 
\end{ruledtabular} 
\end{center} 
\end{table}

Even along the least action trajectory the zero-point vibration energy
remains too high, showing that this kind of parametrization with two
deformation coordinates in which $R_2$ is varied exponentially from an
initial value $R_2=R_0$ to $R_2=R_{2f}$, is not suitable.  The reason is
that the deepest minimum of deformation energy (Fig.~\ref{fl0}), determining
the first turning point of the action integral, is obtained in the
deformation space where the nuclear inertia (see Fig.~\ref{crm}) is too
large. By trying a linearly decreasing law of $R_2$ we haven't got any
better result, as expected.

In principle by using two independent deformation parameters instead of only
one should lead to a final solution closer to reality.  Best results are
obtained for linearly increasing $R_2$.

From our previous experience \cite{p333jpg14}, it seems that by keeping $R_2
= R_{2f}=$~constant we can find a fission trajectory (a given $R_{2f}$ or
$\eta$) along which the reproduction of experimental half-life would be
possible with a reasonable value of $E_v$.  By comparing the optimum values
of zero-point vibration energy from Table~\ref{tab2} (two deformation
parameters with exponential decrease of $R_2$) with those from
Table~\ref{tab3} (one deformation parameter with $R_2=$~constant) it is
clear that the simplest parametrization is more appropriate because $E_v$
(smallest value 1.34~MeV) is about three times smaller than 4.27~MeV.
The detailed potential barriers for ten different light fragments of
fissioning $^{286}$Fl are shown in Fig.~\ref{10b}.
 
Perhaps besides the inappropriate shape parametrization one should also
consider another reason for this discrepancy: the strength parameters of the
spin-orbit {\bf l}{\bf s} and {\bf l}$^2$ terms of the ATCSM are taken to
obtain a proton magic number $Z=114$ --- exactly the case of $^{286}$Fl.

In conclusion, with our method of calculating the spontaneous fission
half-life including macroscopic-microscopic method for deformation energy
based on asymmetric two-center shell model, and the cranking inertia for the
dynamical part, we may find a sequence of several trajectories one of which
gives the least action.  

Assuming spherical shapes, we have tried four laws of variation of the
radius of the light fragment from the initial value at $R=R_i$ to the final
one at the touching point $R=R_t$: exponentially and linearly decreasing,
linearly increasing and $R_2=$~constant.

The shape parametrization with linearly increasing $R_2$ is more suitable
to describe the fission process of SHs in comparison with that of
exponentially or linearly decreasing law. It is in agreement with the
microscopic finding concerning the preformation of a cluster at the surface,
which then penetrates by quantum tunneling the potential barrier.

As far as the potential barrier shape at a given mass asymmetry, there is a
transition from a two hump at lower values to one hump at higher values
around $\eta=0.5$.  The dominant macroscopic component at a high mass
asymmetry, comes from the presence of the Businaro-Gallone mountain.  

The touching point deformation energy versus mass asymmetry shows the three
minima, produced by shell effects, corresponding to three decay modes:
spontaneous fission, cluster decay, and $\alpha$~decay.

All calculations were performed for spherical fragments (the semiaxes ratios
of spheroidally deformed fragments are equal to unity). By considering in
the future the deformed fragments we trust the method could be further
improved.

\begin{acknowledgments} 

This work was supported within the IDEI Programme under Contracts No. 
43/05.10.2011 and 42/05.10.2011 with UEFISCDI, and NUCLEU Programme
PN16420101/2016 Bucharest.  

\end{acknowledgments}


\end{document}